\documentclass[9pt,conference]{IEEEtran}
\IEEEoverridecommandlockouts
\PassOptionsToPackage{dvipsnames}{xcolor}

\usepackage{cite}
\usepackage{amsmath,amssymb,amsfonts}

\usepackage{algorithm}
\usepackage{algorithmic}

\usepackage{graphicx}
\usepackage{textcomp}
\usepackage{xcolor}
\usepackage{lipsum}
\usepackage{CJKutf8}
\usepackage{booktabs}
\usepackage{arydshln}
\usepackage{multirow}
\usepackage{utfsym}
\usepackage{url}
\usepackage{hyperref}
\usepackage{pifont}
\usepackage{listings}
\usepackage{bbding}
\usepackage{pifont}
\usepackage{colortbl}

\def\BibTeX{{\rm B\kern-.05em{\sc i\kern-.025em b}\kern-.08em
    T\kern-.1667em\lower.7ex\hbox{E}\kern-.125emX}}

\begin{document}

\title{Towards Flow-Matching-based TTS \\ without Classifier-Free Guidance
\\
}

\author{
    \IEEEauthorblockN{
        Yuzhe Liang$^{1,2}$, 
        Wenzhe Liu$^3$, 
        Chunyu Qiang$^3$,
        Zhikang Niu$^{1,2}$, 
        Yushen Chen$^{1,2}$, \\
        Ziyang Ma$^1$, 
        Wenxi Chen$^{1,2}$,
        Nan Li$^3$,
        Chen Zhang$^3$, 
        Xie Chen$^{1,2\dag}$\thanks{$^\dag$Corresponding Author.}
    }
    \IEEEauthorblockA{\textit{$^1$MoE Key Lab of Artificial Intelligence, X-LANCE Lab, Shanghai Jiao Tong University, China}}
     \IEEEauthorblockA{$^2$\textit{Shanghai Innovation Institute
}, $^3$\textit{Kuaishou Technology, Beijing, China} 
    } 
}

\maketitle

\begin{abstract}

Flow matching has demonstrated strong generative capabilities and has become a core component in modern Text-to-Speech (TTS) systems. To ensure high-quality speech synthesis, Classifier-Free Guidance (CFG) is widely used during the inference of flow-matching-based TTS models. However, CFG incurs substantial computational cost as it requires two forward passes, which hinders its applicability in real-time scenarios. 
In this paper, we explore removing CFG from flow-matching-based TTS models to improve inference efficiency, while maintaining performance. 
Specifically, we reformulated the flow matching training target to directly approximate the CFG optimization trajectory.
This training method eliminates the need for unconditional model evaluation and guided tuning during inference, effectively cutting the computational overhead in half. Furthermore, It can be seamlessly integrated with existing optimized sampling strategies. We validate our approach using the F5-TTS model on the LibriTTS dataset. Experimental results show that our method achieves a 9$\times$ inference speed-up compared to the baseline F5-TTS, while preserving comparable speech quality. We will release the code and models to support reproducibility and foster further research in this area.

\end{abstract}


\section{Introduction}
\label{sec:intro}

The recent rapid advancements in large language models (LLMs) and diffusion-based generative models have significantly propelled the development of text-to-speech (TTS) synthesis systems. These technological breakthroughs offer new paradigms for generating high-fidelity and expressive speech. Current TTS methods can be categorized into two approaches: autoregressive (AR) \cite {an2024funaudiollm,du2024cosyvoice,wang2023viola} and non-autoregressive (NAR) \cite {ju2024naturalspeech, eskimez2024e2ttsembarrassinglyeasy, chen2024f5, mehta2024matcha}. 

In AR-based models, speech signals are discretized into token sequences and modeled with next-token prediction tasks. These methods capitalize on the inherent in-context learning \cite{brown2020language} capabilities of LLMs to generate high-quality and natural speech. However, AR models often suffer from slow inference speed and error accumulation due to their iterative sequential generation process. On the other hand, NAR-based models have achieved impressive performance, largely driven by recent advances in diffusion-based generative techniques \cite{dhariwal2021diffusion,ho2020denoisingdiffusionprobabilisticmodels,song2022denoisingdiffusionimplicitmodels,nichol2021improveddenoisingdiffusionprobabilistic,lipman2023flow}. 
Notably, flow matching \cite{lipman2023flow}, which has been proven effective in aligning the distribution of generated data with the target distribution through continuous transformations, has emerged as a pivotal technique in both autoregressive and non-autoregressive paradigms, significantly enhancing the speech quality. Although diffusion-based models offer faster synthesis compared to AR models, they still rely on multi-step denoising, which limits their inference efficiency. Recent studies\cite{huang2022fastdiff,huang2022prodiff,jeong2021diff} have explored methods to improve the inference speed of diffusion-based TTS models. ProDiff \cite{huang2022prodiff}, CoMoSpeech \cite{ye2023comospeech}, and Reflow-TTS \cite{guan2024reflow} aim to reduce the number of sampling steps through distillation, while FlashSpeech \cite{ye2024flashspeech} further accelerates sampling via latent consistency modeling.

 Most prior work on optimizing diffusion-based TTS models has focused on reducing the number of sampling steps. However, another core factor affecting inference efficiency is the widespread use of classifier-free guidance (CFG), which requires performing both conditional and unconditional inference for each sampling step, effectively doubling the computational overhead during inference. In the field of image generation, Tang et al.\cite {tang2025diffusionmodelsclassifierfreeguidance} proposed a model-guidance training approach that enables diffusion models to remove the need for CFG at inference time. It is important to note that this method was primarily designed for class-conditional tasks, where the model is trained to generate images based on fixed categories; its application to text-to-speech remains underexplored. Inspired by this pioneering work, in this paper, we explore enhancing speech generation by modifying the flow matching training target, specifically for text-conditioned and audio-conditioned tasks, such as TTS. Our goal is to enable the model to perform only conditional predictions during inference, eliminating the need for unconditional predictions required by traditional CFG. To validate the feasibility of this approach, we adopt F5-TTS, a representative flow-matching-based TTS model, as a case study. Experimental results show that our method can effectively halve the computational cost per sampling step without degrading the generated speech quality compared to the baseline F5-TTS with CFG. Furthermore, our method can be seamlessly combined with advanced sampling strategies to achieve additional speedup, highlighting its potential for real-time applications.

Our contributions are summarized as follows:

\begin{itemize}
    \item 

    We present the first attempt to remove classifier-free guidance from flow-matching-based TTS models at inference time by adopting the model-guidance
training to alter the prediction target of flow matching. 

    \item 

    We validate our approach on F5-TTS, effectively halving the inference cost without compromising the quality of the generated speech. Moreover, the proposed method can be seamlessly integrated with existing optimized sampling strategies, resulting in further speedup.

\end{itemize}

\section{Related work}

In this section, we provide a brief overview of flow matching \cite{lipman2023flow}, classifier-free guidance \cite{ho2022classifier}, along with our study case F5-TTS \cite{chen2024f5}.

\subsection{Flow Matching}

Flow Matching (FM) \cite{lipman2023flow} introduces a novel approach to train continuous normalizing flows (CNFs) by directly optimizing the congruence between the learned and target probability paths. This method typically constructs a closed-form solution based on optimal transport (OT) theory, eliminating the need for iterative score matching across discrete timesteps as required in conventional diffusion models \cite{songscore}. (OT-)FM \cite{tong2023improving} leverages a linear interpolation path between the prior noise \( z \sim \mathcal{N}(0, I) \) and data samples \( x_1 \sim q(x) \), defined as \( x_t = (1 - t) \cdot z + t \cdot x_1 \) for \( t \in [0,1] \). The training target focuses on minimizing the discrepancy between the learned vector field \( v_\theta \) and the theoretical optimal transport dynamics, formulated as
\begin{equation}
    \mathcal{L}_{\text{FM}} = \mathbb{E}_{t \sim \mathcal{U}(0,1), z, x_1} \left\| v_\theta(x_t, t) - (x_1 - z) \right\|^2.
\end{equation}
During synthesis, samples are generated by solving the ordinary differential equation
\begin{equation}
    \frac{dx}{dt} = v_\theta(x_t, t)
\end{equation}
with the initial condition \( x(1) \sim p_0(z) \) evolving to \( x(0) \sim q(x) \). This framework not only maintains the theoretical guarantees from optimal transport but also achieves superior computational efficiency compared to conventional diffusion approaches.

In NAR models like Voicebox \cite {le2023voicebox}, E2 TTS \cite {eskimez2024e2ttsembarrassinglyeasy}, and F5-TTS \cite {chen2024f5}, flow matching is employed to predict the mel spectrogram based on text conditions. In the field of AR models like CosyVoice \cite {du2024cosyvoice}, Seed-TTS\cite{seedtts}, and FireredTTS \cite{guo2024fireredtts}, semantic tokens generated by a language model (LM) and auxiliary information like speaker embeddings are used as conditions for the second-stage mel spectrogram prediction. Unlike CosyVoice, which first generates discrete tokens using a LM, ARDiT \cite {liu2024autoregressive} and DiTAR \cite {ditar} directly receive latents, and autoregressively generate continuous tokens.  

\subsection{Classifier-Free Guidance}
Classifier-Free Guidance (CFG)\cite{ho2022classifier} is a pivotal advancement in conditional diffusion modeling, designed to enhance the alignment between generated samples and conditioning signals without relying on auxiliary classifiers. This technique has been widely adopted in various domains, including text-to-speech synthesis and cross-modal generation\cite{wang2024frieren}. The core mechanism involves interpolating predictions from both conditional and unconditional diffusion processes during sampling. Specifically, the guided noise prediction is formulated as:

\begin{equation}
    \tilde{\epsilon}_\theta(x_t, t, c) = \epsilon_\theta(x_t, t) + \omega \cdot \left( \epsilon_\theta(x_t, t, c) - \epsilon_\theta(x_t, t) \right)
\end{equation}
where $\omega \geq 0$ denotes the guidance scale. When $\omega = 0$, the generation degenerates to an unconditional process governed by $p_\theta(x)$; increasing $\omega$ amplifies the conditioning effect, thereby improving semantic consistency with the input $c$ at the potential cost of sample diversity. Empirical studies suggest that an optimal $\omega$ value balances artifact suppression and conditional fidelity.

Despite the remarkable success of CFG in diffusion modeling, it still faces many challenges in practical applications. For example, a high guidance scale tends to trigger mode collapse\cite {mokady2022nulltextinversioneditingreal,chung2024cfgmanifoldconstrainedclassifierfree,fan2025cfgzeroimprovedclassifierfreeguidance}. More critically, although CFG performs well in practice, its working principle has not been fully explained theoretically \cite {bradley2024classifierfreeguidancepredictorcorrector}.

\subsection{F5-TTS}
The F5-TTS \cite{chen2024f5} model is based on flow matching with Diffusion Transformer (DiT), which effectively avoids complex designs such as duration models, text encoders, and phoneme alignment by directly populating the character sequences and employing a simple text-guided speech filling task for training. In addition, F5-TTS introduces a Sway Sampling strategy for inference, which significantly improves model performance and efficiency, and can be easily applied to other flow-matching-based models without retraining. While maintaining a concise text-to-speech generation process, F5-TTS demonstrates highly natural, smooth speech synthesis and seamless code-switch in zero-shot generation tasks, a significant improvement over other diffusion model TTS models.

\section{Method}

\subsection{Limitations of Classifier-Free Guidance in Flow Matching}

The application of Classifier-Free Guidance (CFG) to flow matching reveals inherent theoretical inconsistencies between training and inference. To formalize this, we first revisit the CFG's theoretical foundation in diffusion models. Let $p(x|y)$ denote the conditional data distribution and $p(x)$ the unconditional counterpart. From the scope of score matching, the conditional gradient can be decomposed via Bayes' rule:
\begin{equation}
    \nabla_x \log p(x|y) = \nabla_x \log p(x) + \underbrace{\nabla_x \log p(y|x)}_{\text{Implicit Classifier}}
    \label{eq:bayes_decomp}
\end{equation}
where the term $\nabla_x \log p(y|x)$ acts as an implicit classifier \cite{dhariwal2021diffusion}. CFG amplifies this classifier by introducing a guidance scale $w$:
\begin{equation}
    \nabla_x \log \hat{p}(x|y) = \nabla_x \log p(x) + w \cdot \nabla_x \log p(y|x).
    \label{eq:cfg_score}
\end{equation}

Reformulating \eqref{eq:cfg_score} using \eqref{eq:bayes_decomp}, we derive the CFG-adjusted gradient:
\begin{equation}
    \nabla_x \log \hat{p}(x|y) = (1 - w) \cdot \nabla_x \log p(x) + w \cdot \nabla_x \log p(x|y)
    \label{eq:cfg_grad}
\end{equation}
establishing a linear interpolation between unconditional and conditional dynamics.

\paragraph{CFG in Flow Matching} extends this concept to vector fields. Let $u_t(x)$ denote the marginal velocity field and $u_t(x|y)$ the conditional counterpart. The CFG-adjusted field becomes:
\begin{equation}
    v_t^{\text{CFG}}(x|y) = u_t(x) + w \cdot \left( u_t(x|y) - u_t(x) \right).
    \label{eq:cfg_velocity}
\end{equation}

However, a critical limitation arises from the training-inference mismatch. During training, the model jointly estimates both fields through conditional dropout:
\begin{equation}
    v_t^{\text{train}}(x|y) = 
    \begin{cases}
        u_t(x|y), & \text{w.p. } p_{\text{cond}} \\
        u_t(x),   & \text{otherwise}.
    \end{cases}
    \label{eq:training_target}
\end{equation}

Whereas inference applies the interpolated field \eqref{eq:cfg_velocity}, this discrepancy induces two key issues. The shared network parameters must simultaneously satisfy conflicting targets for \( u_t(x) \) and \( u_t(x|y) \), which causes parameter conflict. Additionally, the unconditional estimate \( u_t(x) \) becomes biased toward conditional trajectories due to the interpolation term. These challenges highlight the need to address the inherent trade-offs between unconditional and conditional targets in the model's training and inference processes.

\subsection{Flow Matching without Classifier-free guidance}

\begin{algorithm}[t]
   \caption{Training Conditional-Flow-Matching-based TTS with Model-Guidance}
   \label{alg:training-tc}
    \begin{algorithmic}[1]
       \STATE {\bfseries Input:} dataset $\{\mathbf{X_i},\mathbf{C_i}\}$, model parameters $\theta$, $\phi$, scalar $s$
       \REPEAT
       \STATE Sample data $(x_0, c)\sim\{\mathbf{X_i},\mathbf{C_i}\}$
       \STATE Sample time $t\sim\mathbf{U}(0,1)$
       \STATE Compute change in velocity field with gradient stopping:
       
       $\Delta v_t(x_0, c) = \text{sg} (v_t(x_0, c) - v_t(x_0))$
       
       \STATE Modify target velocity field:
       
        $u_t'(x_0 | c) = u_t(x_0 | c) - w \cdot \Delta v_t(x_0, c)$
       
       \STATE Compute loss:
       
       $\mathcal{L}_{\text{MG-CFM}} = \| v_t(x_0)  - u_t'(x_0 | c) \|_2^2$
       
       \STATE Back propagation:
       $
       \theta = \theta - \eta \nabla_\theta \mathcal{L}_{\text{MG-CFM}}
       $
       \STATE Update model parameters
       \UNTIL{converged}
    \end{algorithmic}
\end{algorithm}

According to Tang et al.\cite{tang2025diffusionmodelsclassifierfreeguidance}, in the field of image generation base on diffusion model, it has been demonstrated that directly incorporating the objective of CFG into training, by enabling the model to fit the linear outcomes of conditional and unconditional predictions, can significantly enhances the model’s capabilities and reduces inference overhead by a factor of two.

Flow matching has proven effective by transforming noise into target data. To enhance this, we modify the predicted vector field to introduce a discrepancy between the conditional and unconditional velocities. The loss function of conditional flow matching is defined as:

\begin{equation}
\mathcal{L}_{\text{CFM}} = \mathbb{E}_{t, p_t(x| y), p(y)} \| v_t(x| y) - u_t(x | y) \|_2^2 .
\label{eq:cfm loss}
\end{equation}

By referring to Equation \eqref{eq:cfg_velocity}, we can substitute the expression for $v_t^{{\text{CFG}}}(x| y)$ into the loss function to obtain a target changed loss function:
\begin{equation}
\mathcal{L}_{\text{MG-CFM}} = \mathbb{E}_{t, p_t(x \mid y), p(y)}  \| v_t(x| y) + w \cdot \Delta v_t(x, y) - u_t(x | y) \|_2^2 
\label{eq:cfm_loss2}
\end{equation}
where \(\Delta v_t(x| y) =sg(v_t(x| y) - v_t(x))\), and $w$ is a parameter that controls the weight of the difference term. Here, ``\( sg \)" indicates the stop gradient operation, ensuring the parameters of the unconditional model remain unaffected during the computation of the conditional change.

\section{Experiments}

\subsection{Datasets}

For training, we employed the LibriTTS dataset\cite{zen2019libritts}, which is a multi-speaker English corpus comprising approximately 585 hours of read speech from 2,456 native English speakers, sampled at 24 kHz.

For evaluation, we followed the F5-TTS \cite{chen2024f5} evaluation setup and utilized the LibriSpeech-PC subset as the test set to comprehensively assess the model’s performance in English. This subset comprises 1,127 samples and contains 2.2 hours of speech from the LibriSpeech test-clean set.

\subsection{Model and training configurations}

We follow the F5-TTS setup and use the base model as our baseline, which consists of 22 layers, 16 attention heads, and a 1024-dimensional embedding for the DiT block; and 4 layers with a 512-dimensional embedding for the ConvNeXt V2 module, totaling 335.8M parameters. All models are trained for 500K updates with a frame-level batch size of 307,200 audio frames (approximately 0.91 hours of audio data) using 8 NVIDIA A800 80G GPUs.

The AdamW optimizer is employed with a peak learning rate of \(7.5e {-5}\), which is linearly warmed up for the first 20k updates and then linearly decayed for the remainder of the training period. We set the maximum gradient norm clipping to 1 to ensure stable training.

In the training process, we set the parameter \(w\) to 0.7. This parameter controls the influence of conditional information on the velocity field, allowing for a balanced integration of unconditional and conditional during training. We kept the previous CFG training in the same way, with the probability \(p_{uncond}\) dropping all the conditions, to ensure the fairness of the experiment. Specifically, \(p_{uncond}\) is set to 0.2, and we independently increase the chance of dropping the audio condition to 0.3. These probabilities are set independently of each other.

Audio samples are represented by 100-dimensional log mel-filterbank features, sampled at 24 kHz with a hop length of 256. During training, 70\% to 100\% of mel frames are randomly masked for the infilling task.

\subsection{Evaluation Metrics}

We evaluate model performance using a cross-sentence task, where the model synthesizes speech from a reference text, guided by a short speech prompt and its transcription. Objective metrics include Word Error Rate (WER) and speaker similarity (SIM-O). WER is calculated using Whisper-large-v3. SIM-O measures cosine similarity between synthesized and target speeches using a WavLM-large-based \cite{wavlm} speaker verification model. To evaluate the generated speech quality from a listening perspective. we used the NISQA model \cite{mittag2021nisqa}, which was trained on extensive subjective listening test data via deep learning, to measure the Mean Opinion Score (MOS), a standard subjective assessment where evaluators rate speech samples on a five-point scale. 
Additionally, we assess the time efficiency of the speech synthesis system using the Real-Time Factor (RTF), which quantifies the time required to generate one second of speech.  Specifically, we used 3s of speech as the prompt, generated 10s of speech, and repeated one 100 times to get the average RTF on  NVIDIA GeForce RTX 3090.

\subsection{Evaluation Results}

In Table \ref{tab:libritts_exp},  when inference with CFG, the guidance scale is set to 2. Based on the results, we know that flow matching without CFG allows the model to achieve slightly better than F5-TTS in terms of speaker similarity and significantly better than F5-TTS in terms of WER and MOS. Notably, this is accomplished with only half of the inference overhead required by previous methods.

Furthermore, we observed that when adopting the new training approach, using the previous CFG inference method with 32 sampling steps during the inference phase still achieves comparable improvements in WER and speaker similarity (SIM). This indicates that this training method effectively addresses the training-inference mismatch in this specific scenario. It also paves the way for a deeper exploration of the underlying mechanisms of CFG in future research.

\subsubsection{Training Expense}

Our method introduces additional model forward passes during the training phase to explicitly differentiate between predictions with and without conditions. While this approach inevitably increases training overhead, we employ two forward passes and a single backpropagation step. As a result, the overall training overhead is approximately 1.5 times higher than that of the previous methods.

\begin{figure}[t]
  \centering
  \includegraphics[width=8cm]{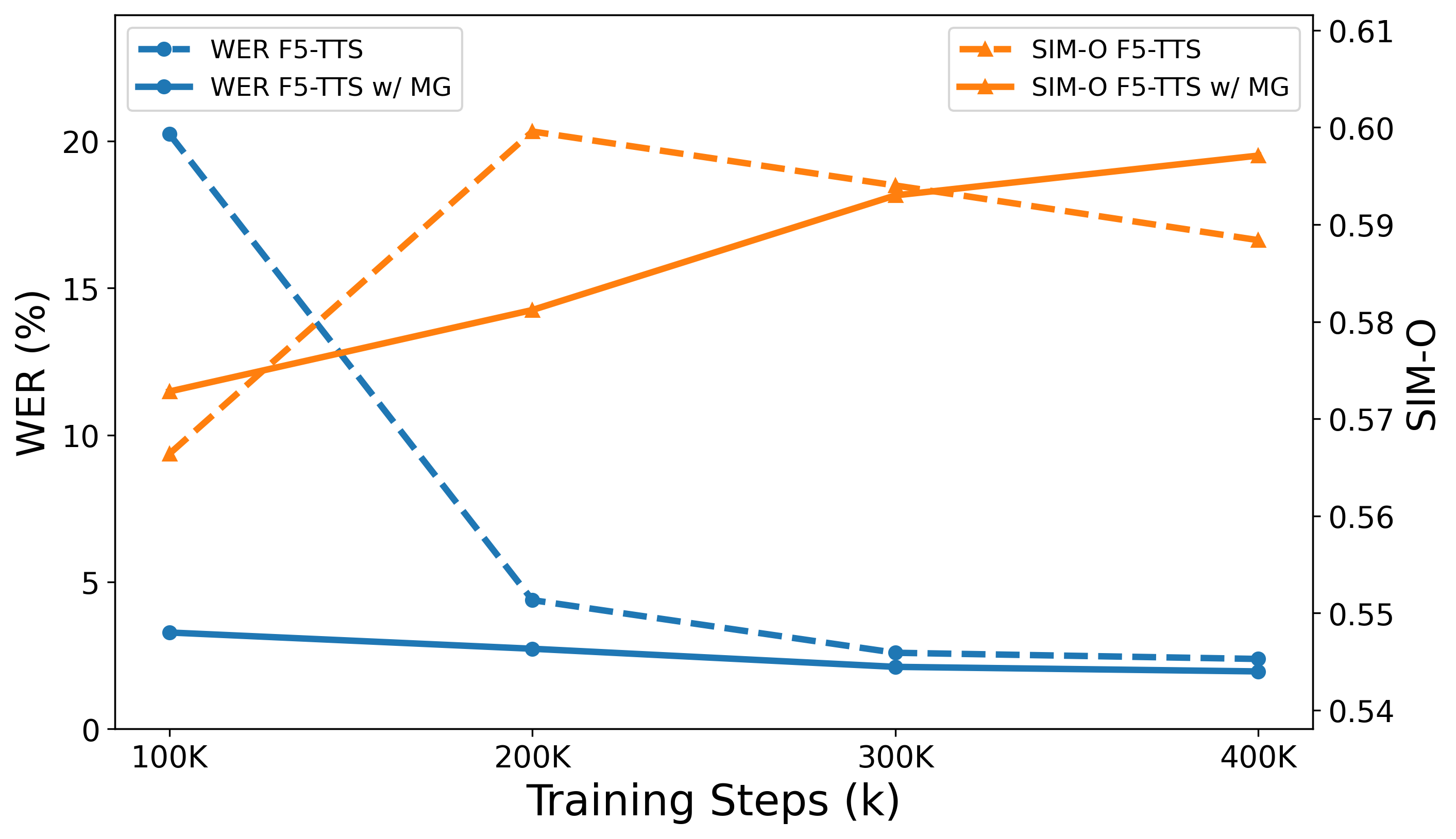}
  \caption{Comparison between the baseline F5-TTS and our proposed approach at different training steps. Solid lines represent the WER and SIM-O metrics across training steps for the proposed F5-TTS trained with the model-guidance method, while dotted lines correspond to the baseline F5-TTS results. The results show that training flow-matching models without CFG leads to more stable training and faster convergence.}
  \label{fig:traing step}
  \vspace{-2mm}
\end{figure}

Meanwhile, we observe that our proposed training target significantly accelerates convergence, achieving comparable results with approximately half the training steps needed by F5-TTS. As shown in Fig. \ref{fig:traing step}, the result for F5-TTS in the figure uses CFG inference, doubling the sample steps. Our method converges very quickly, especially in terms of WER. When F5-TTS has not yet converged at 100k and 200k steps, our method has already reached a lower value.

To summarize, this method does not increase training overall overhead and is even better than the previous method in terms of training time.

\subsubsection{Ablations}

In the ablation experiments, we mainly experimented with the effects of different \(w\) parameters and whether or not to use gradient stops in the experiments.

\begin{figure}[t]
  \centering
  \includegraphics[width=8cm]{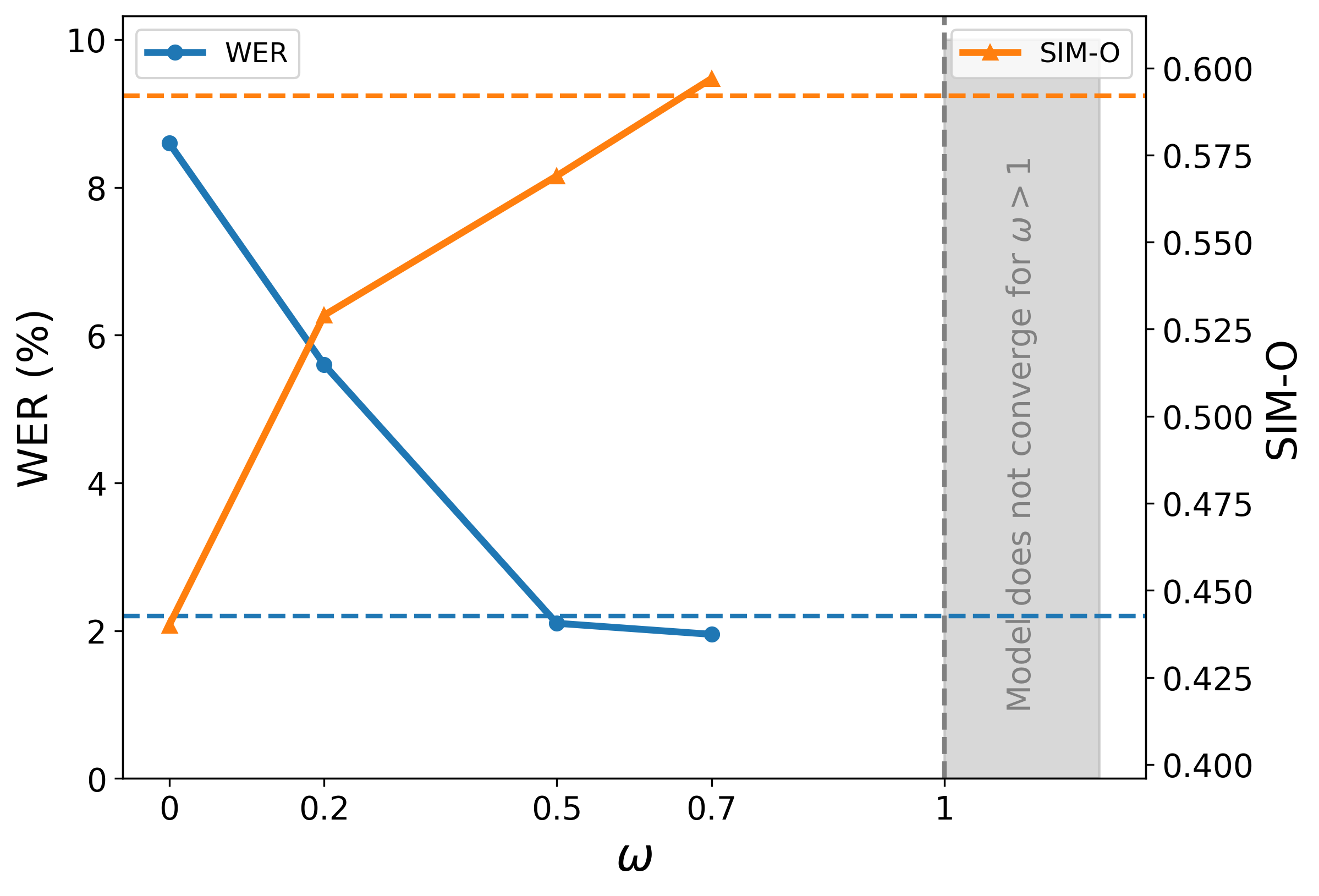}
  \caption{Effect of the hyperparameter \(w\)  (as defined in Equation (\ref{eq:training_target})) during training. The orange dotted line and blue dotted line represent the SIM-O and WER of the baseline F5-TTS inference, respectively. We observe that at \(w\)  = 0.5, the proposed approach outperforms the baseline F5-TTS in terms of WER, and at \(w\)  = 0.7, it achieves improvements in both WER and SIM-O compared to baseline F5-TTS.}
  \label{fig:w_ablation}
  \vspace{-2mm}
\end{figure}

\begin{itemize}
    \item Effect of different $w$ parameters:
    During the experimental process, we discovered that the model is highly sensitive to the value of \( w \). When \( w \) is slightly increased, model training tends to collapse. Based on empirical values from Classifier-Free Guidance (CFG) reasoning, we aimed to set \( w \) close to the target value of 2, which is used in the F5-TTS inference process. However, during actual training, we found that setting \( w \geq 1 \) led to highly unstable training, with significant fluctuations in the loss.

    The results shown in Fig. \ref{fig:w_ablation} correspond to the model's inference conducted without employing CFG. We observed that the model's performance improves with increasing \( w \), which aligns with the goal of CFG inference. The gray area in the figure indicates that the model training fails due to gradient explosion. In terms of the Word Error Rate (WER) metric, single-step inference outperforms the original CFG inference when \( w \) reaches 0.5. At \( w = 0.7 \), single-step inference already surpasses the original CFG inference in both the similarity (SIM-O) and WER metrics.

    Regarding the explanation for the model's collapse when \( w \) is too large, the training target term \( w \cdot (v_t(x_0, c) - v_t(x_0)) \) may become more volatile, thereby disrupting normal training.

    \item Use of stop gradient:
    we found that if we don't use stop gradient for training, the model training loss will not collapse, but it will lead to the model can't inference normally, and the inference results are all noise, the analysis is that if we don't use stop gradient, the \(v_t(x_0)\) gradient back-propagation will cause the model to learn the value of a trival.

\end{itemize}

\begin{table}[htbp]
\centering
\caption{
Performance comparison between the baseline F5-TTS model (FM w/ CFG) and our proposed approach (FM w/o CFG) on the LibriSpeech-PC test set. ``CFG in infer'' indicates whether or not classifier-free guidance is used during inference. The gray background indicates the vanilla F5-TTS inference method, which serves as the baseline.
}
\label{tab:libritts_exp}
\resizebox{1\linewidth}{!}{
\begin{tabular}{lccccccc}
\toprule
\multirow{2.5}{*}{\textbf{Training}} & \multirow{2.5}{*}{\textbf{CFG in infer}} & \multirow{2.5}{*}{\textbf{NFE}}  & \multicolumn{4}{c}{\textbf{LibriSpeech-PC} } \\ 
\cmidrule(lr){4-7}
& & & SIM-O ↑ &  WER(\%) ↓ & RTF ↓ & MOS ↑  \\ 
\midrule
\multirow{6}{*}{FM w/ CFG}
 & \Checkmark (baseline) & \multirow{2}{*}{32} & \cellcolor[HTML]{C0C0C0}0.592 & \cellcolor[HTML]{C0C0C0}2.28 & \cellcolor[HTML]{C0C0C0}0.31 & \cellcolor[HTML]{C0C0C0}4.026 \\
 & {\ding{55}} &  & 0.447 & 8.61 & 0.17 & 3.609 \\
 \cline{2-7}
 & \Checkmark & \multirow{2}{*}{16} & 0.588 & 2.37 & 0.15 & 3.974 \\
 & {\ding{55}} &  & 0.441 & 9.12 & 0.09 & 3.380 \\
 \cline{2-7}
 & \Checkmark & \multirow{2}{*}{7} & 0.582 & 2.46 & 0.07 & 3.799 \\
 & {\ding{55}} &  & 0.457 & 8.43 & 0.04 & 2.641 \\
\hline
\multirow{6}{*}{FM w/o CFG} 
 & \Checkmark & \multirow{2}{*}{32} & 0.651 & 1.80 & 0.31 & 3.939 \\
 & {\ding{55}} &  & 0.597 & \textbf{1.96} & 0.17 & \textbf{4.159}  \\
 \cline{2-7}
 & \Checkmark & \multirow{2}{*}{16} & 0.610 &  3.42 &  0.15 & 3.682 \\
 & {\ding{55}} &  & 0.596 & 2.05 & 0.09 & 4.132 \\
 \cline{2-7}
 & \Checkmark & \multirow{2}{*}{7} & 0.523 &  9.94 & 0.07 &  2.878 \\
 & {\ding{55}} &  & \textbf{0.601} & 2.02 & \textbf{0.04} & 4.101 \\
\bottomrule
\end{tabular}
}
\vspace{-4mm}
\end{table}

\section{Conclusion}

Given the rising adoption of flow matching in modern text-to-speech (TTS) models, optimizing the inference efficiency of these models has become increasingly important, as they typically require multiple sampling steps to achieve satisfactory performance. While most prior work has focused on reducing the number of sampling steps, another critical issue lies in the use of Classifier-Free Guidance (CFG), which requires two forward passes and introduces additional computational overhead. In this paper, we address this issue by exploring how to remove CFG from flow-matching-based TTS models, enabling a single forward pass per sampling step. By reformulating the flow-matching training objective to directly approximate the output of CFG inference, we eliminate the need for unconditional predictions and guidance scale adjustments during inference. This approach significantly reduces computational cost and simplifies the inference procedure. Using F5-TTS as a case study, we demonstrate that inference cost can be effectively halved without sacrificing speech quality. In future work, we plan to extend this method to other flow-matching-based TTS models and evaluate its effectiveness on larger datasets.

\bibliographystyle{IEEEtran} 
\bibliography{refs}

\end{document}